# Orion Routing Protocol for Delay Tolerant Networks

Samir MEDJIAH and Toufik AHMED

CNRS-LaBRI, University of Bordeaux-1.
351 Cours de la Libération,
33405 Talence – France

{medjiah, tad}@labri.fr

*Abstract*—In this paper, we address the problem of efficient routing in delay tolerant network. We propose a new routing protocol dubbed as ORION. In ORION, only a single copy of a data packet is kept in the network and transmitted, contact by contact, towards the destination. The aim of the ORION routing protocol is twofold: on one hand, it enhances the delivery ratio in networks where an end-to-end path does not necessarily exist, and on the other hand, it minimizes the routing delay and the network overhead to achieve better performance. In ORION, nodes are aware of their neighborhood by the mean of actual and statistical estimation of new contacts. ORION makes use of autoregressive moving average (ARMA) stochastic processes for best contact prediction and geographical coordinates for optimal greedy data packet forwarding. Simulation results have demonstrated that ORION outperforms other existing DTN routing protocols such as PRoPHET in terms of end-to-end delay, packet delivery ratio, hop count and first packet arrival.

## I. Introduction

Delay/Disruption Tolerant Network (DTN) may often refer to sparse mobile ad hoc network, where an end-to-end routing path does not necessarily exist. In DTNs, both nodes and links may be inherently unreliable. Due to these constraints, these networks are referred to as *"challenged networks"* [1]. Many other emerging communication networks fall into this paradigm. Vehicular ad hoc networks (VANETs), mobile sensor networks, and nomadic community networks are few examples.

An interesting DTN example is the city bus network, in which nodes consist of buses (cars, taxis, trams…) and communicate using short-range radios. With this type of networks, we can envision a lot of new applications: urban sensing, information dissemination (advertisement, traffic information, buses software update…) or even Internet access. The proper functioning of such applications relies essentially on the efficiency of the routing task. However many challenges affect the routing in DTNs such as the changing network topology due to intermittent connectivity which is inherent to mobile networks as well as to static networks (in the case of low duty cycle of the nodes), and it results in low delivery ratio and high end-to-end delay. The problem of intermittent connectivity can be mitigated if the exact schedule or the dynamics of the network is known in advance. However, this is not often the case in DTNs as building this knowledge is an important issue. Thus, the efficiency of a DTN routing protocol relies essentially on the amount of network knowledge or *"oracles"* (information about contacts, queues or even data traffic) available to perform routing decisions.

In this paper, we propose ORION, a routing protocol for mobile DTNs that capitalizes on the localization information of the nodes (geo-coordinates) and the nature of contacts between this type of nodes (buses, cars, taxis, trams) in an urban area. The contribution presented in this paper is twofold. First, we have investigated the inter-nodes encounter behavior. Second, based on this behavior analysis, we proposed ORION, a novel routing protocol that relies on predicting future contacts between nodes and greedy geographic forwarding of data packets. Thus, with ORION protocol, a communicating node will incrementally build knowledge about its network regarding the inter-nodes encounters behavior and nodes positions. Thereby, it should be able to predict when it will be in contact with other nodes and for how long (duration).

The remainder of this paper is organized as follows. Section II presents the state of the art for DTN protocols and the use of stochastic processes and time series analysis in network communication modeling. Section III presents the ORION protocol. Section IV provides the simulation results and related discussion. Finally, section V concludes the paper.

## II. Related Work

Several routing protocols have been proposed for DTNs. These protocols can be classified into two categories; replication-based and prediction (forwarding)-based protocols. With replication-based protocols, the contacts are assumed to be totally opportunistic and the required topology knowledge at each node is minimal. In this case, the simplest way to deliver a message is to send a copy to each encountered node. This is repeated until the destination receives the message. The Epidemic Routing protocol [2] envisions this strategy. With prediction-based protocols, only a single copy exists across the network at a given time. The protocol needs to be supplied with more knowledge about the network. Given the unavailability of topology information, some protocols try to use probabilities to predict the contact. However, such prediction can be at the price of reduced delivery ratio. Most of the existing prediction-based routing protocols focus mainly on whether two nodes would be in contact in the future, without paying much attention to "when" the contact will happen or "for how long" the contact will last. This lack of contact timing information degrades the contact prediction accuracy and negatively impacts the routing performance.

### A. DTN Routing Protocols Taxonomy

As mentioned earlier, the replication-based routing strategy can achieve high delivery ratio while operating with minimal knowledge. However, this strategy is not optimal in terms of transmission and buffer size. It also suffers from the lack of scalability. Some protocols, adopting this strategy cope with this problem by bounding the number of copies in

the network trading delay for buffer occupancy. To limit the replication, two solutions are used:
- Fix the number of copies and spread them through distinct nodes. *Spray & Wait* routing protocol [3] uses this solution, also called *quota-based solution*.
- Use metrics based on historical encounters between nodes to decide whether to send a copy or not. PRoPHET [4] (Probabilistic Routing Protocol using History of Encounters and Transitivity) protocol uses this solution.

The PRoPHET protocol utilizes an algorithm that makes use of the non-random aspect of the real world. This is done by maintaining a set of delivery success probabilities to known destinations, and by replicating messages during opportunistic contacts. Replication is done only for an encountered node which does not have a copy of the message and has a good probability to deliver the message to its final destination. Given a node *i*, the probability of node *i* to encounter another node *j* is denoted as *P(i,j)*. The delivery probabilities are computed during each contact driven by the following three rules:

a) *Updating:*
$$P(i,j)_{new} = P(i,j)_{old} + (1 - P(i,j)_{old}) \times L_{encounter}$$

Where $L_{encounter}$ is an initializing constant.

b) *Aging:* $P(i,j)_{new} = P(i,j)_{old} \times \gamma^n$

Where $\gamma$ is an aging constant and $n$ denotes the number of time units elapsed since the last aging.

c) *Transitivity:*
$$P(i,k)_{new} = P(i,k)_{old} + (1 - P(i,k)_{old}) \times P(i,j) \times P(j,k) \times \beta$$

Where $\beta$ is a scaling constant.

In the prediction (forwarding) based protocols, a node is associated with a forwarding quality/probability metric for each destination, which is usually a direct (*one*-hop) forwarding quality such as contact frequency [6], or time elapsed since last contact [7][8][9].

During a contact, if a node *i* encounters another node *j*, node *i* will decide whether to send the message to node *j* based on the comparison between the direct forwarding qualities of node *i* and node *j*. The main drawback of this approach lies in the fact that good forwarding is not guaranteed due to these observations:
1. Node *j* with a better forwarding quality than node *i* does not necessary mean that node *j* is a good forwarder.
2. Although the quality of node *j* is high, node *i* may encounter better nodes in the near future.
3. Similarly, even though the forwarding quality of node *j* is lower than node *i*, node *j* may be still the best forwarder that node *i* could encounter in the future.

### B. Times Series in Network Modeling

The proposed ORION protocol makes use of time series to predict contacts. A time series is an ordered sequence of values of a variable $\{y_t\}_{t \in T}$ indexed by an ordered set $T = \{t_1, t_2, t_3, ..., t_n\}$. The time series analysis serves two purposes: (1) Obtain an understanding of the underlying forces and structure that have produced the observed data, and (2) Fit a model and proceed to forecasting, monitoring or even feedback and feedforward control. Time series analysis is used for many applications such as economic forecasting, sales forecasting, budgetary analysis, stock market analysis, yield projections, process and quality control, etc. Recently, it starts being used in the field of computer networks communications. Indeed, time series have gained the attention of many researchers for the modeling of the Internet and wireless mobile networks traffic. In [10], Basu *et. al.* have modeled the Internet traffic using ARMA process of order *(p,q)*. Using this model, they predict the traffic generated by a TCP source using FDDI protocol. In [11], Liu *et. al.* have proposed an energy efficient technique for data collection in Wireless Sensor Networks. A sensor is hold from transmitting redundant data. The data are not sent if they can be predicted by the sink node. For prediction, they utilize ARIMA model of order *(p,d,q)* [12] due to its outstanding model fit and small computational cost. In [13], Herbert *et. al.* extend this idea to the hierarchic routing protocol LEACH [14] by providing verification at the cluster head. This approach has shown great communication cost savings. In [15], Banerjee *et. al.* used a *birth and death* process to model the network's dynamics. A node entering in the transmission range of a source node is considered as a *birth*. Similarly, a *death* refers to when it leaves this range. Finally, in [16], Singh *et. al.* extend this idea by using an AutoRegressive (AR) process to model the number of a node's neighbors in a mobile ad hoc network.

When dealing with stochastic processes, values of the involved random variables are taken over time forming the time series for further analysis. An important step while analyzing time series is to determine the suitable model (or class of models) fitting the observed data. A common approach to analyze time series is the use of *ARMA* (AutoRegressive Moving Average) analysis since it can be used for stationary and non-stationary processes. An *ARMA* process is a combination of an Autoregressive process (*AR*) and a Moving Average (*MA*) process. In an *AR* process, a random variable is "explained" by its past values rather than other variables. While with *MA* process, a random variable is supposed to be explained by its actual mean, augmented by a weighted sum of the errors (random shocks) that tainted the previous values. *ARMA* analysis was introduced by Box and Jenkins [17] and they have identified three steps to model and forecast time series:
1. *Model Identification:* this step is performed to estimate a model structure by using two essential functions: the autocorrelation function (ACF) and the partial autocorrelation function (PACF).
2. *Parameter Estimation*: this step is performed for fitting the identified model to the observed data. This is achieved by determining the coefficients of the linear combination.
3. *Forecasting*: the final objective is to predict the future values of the time series based on the already observed data and the linear combination estimated at the second step.

And so, $ARMA(p,q)$ model is defined as:

$$y_t = c + \mu + \sum_{i=1}^{p} \varphi_i y_{t-i} + \sum_{j=1}^{q} \theta_j \varepsilon_{t-j} + e_t \quad (Eq.1)$$

Where:
$p$ Non-negative integer; order of the process *AR*,
$q$ Non-negative integer; order of the process *MA*,
$\varphi_i$ Time-invariant coefficients of the *AR* model,
$\theta_i$ Time-invariant coefficients of the *MA* model,
$e_t$ Samples of white noise with mean zero and variance $\sigma^2$,
$\varepsilon_t$ White noise error terms,

*c*  A constant (often omitted),
*μ*  Expectation of *y* (often assumed to be equal to zero)

To be considered for ARMA analysis, a time series must be stationary. To verify the stationarity two conditions must hold:

$E(y_t) = \mu$ is constant independent of instant *t*.  (Eq.2)

$Cov(y_t, y_{t-j}) = \gamma^j$ only depends on time lag *j*.  (Eq.3)

## III. ORION ROUTING PROTOCOL

### A. Target Application

In this paper, we have considered a city-bus network in which the communicating nodes consist of buses, trams, cars and hotspots. The buses and trams are assumed to be "*regular*" mobile nodes, where the cars are assumed to be "*random*" mobile nodes and finally the hotspots and access points to be fixed nodes. The regular nodes move across the area along a certain trajectory, while random nodes move freely across the urban area. In this scenario, all mobile nodes move with a non-constant speed. Each communicating node is assumed to be equipped with localization hardware. Consequently, we propose to utilize geographic addressing and achieve data packets forwarding in a greedy fashion based on distance and/or angle calculus.

Following the Box and Jenkins steps to model the times series using ARMA model, we have conducted some simulations of our city-bus network. In these simulations, instead of using synthetic mobility models [18][19][20], we studied the two time series in pseudo-realistic environment mapped on a real city map, namely Bordeaux in France. Additional information concerning contact analysis and the simulations conducted can be found at [21].

ORION is based on greedy geographic forwarding and contact prediction. In the following, we explain our methodology to achieve an efficient contact prediction.

### B. Contact Behavior Analysis

For our analysis and in order to use efficiently time series, we propose to discrete the time into small periods of time $\Delta t$. In the description, we further denote $(1\Delta t, 2\Delta t, ..., n\Delta t)$ as the time instants $(t_1, t_2, ..., t_n)$.

In networks with intermittent connectivity, a node becomes aware of an eventual contact by the mean of periodically exchanged *HELLO* messages. Consequently, contact (connection) duration *C* with a certain node is the sum of consecutive periods of time $\Delta t$ over which the node received at least one *HELLO* message from the other node. Respectively, the duration of the non-contact (disconnection) $\bar{C}$ is the sum of consecutive periods of time $\Delta t$ over which the node did not receive any *HELLO* messages from the other node. The duration of a contact *C* and a non-contact $\bar{C}$ are two random variables.

In order to study the contact behavior and based on the consecutive values of the two random variables *C* and $\bar{C}$, we construct the two times series $\{C_t\}_{t \in N}$ and $\{\bar{C}_t\}_{t \in N}$, where $C_i$ denotes the duration of the $i^{th}$ contact (connection), and $\bar{C}_i$ represents the duration of the $i^{th}$ non-contact (disconnection). **N** is the set of natural integers.

Examples of $\{C_t\}_{t \in N}$ and $\{\bar{C}_t\}_{t \in N}$ chronograms are shown in Figure 1. To apply the Box-and-Jenkins approach, we had to verify the stationarity of the two time series. Thus, we run the stationarity test (see *Eq.2* and *Eq.3*). The results showed that the two stationarity conditions hold for almost all the time series obtained from the simulation (at the rate of two times series $C_i$ and $\bar{C}_i$ by contacted node at each node). Consequently, $C_i$ and $\bar{C}_i$ can be analyzed using ARMA analysis.

Based on this information, a node can predict the future value of the contact's duration (connection's duration), and also the future value of the non-contact duration (disconnections' duration). Consequently, the node will be able to predict when the next contact will happen and for how long it will last. This knowledge will be extremely beneficial to perform routing decisions.

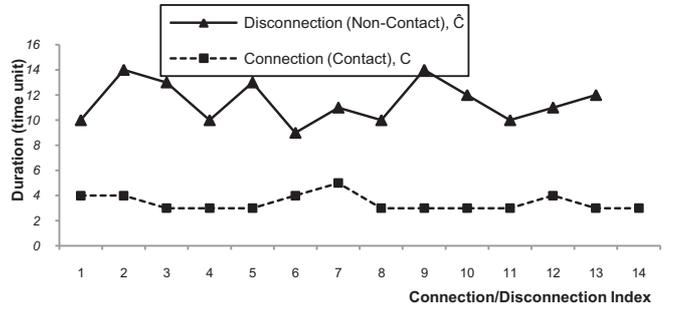

Figure 1: Variation of $C_i$ and $\bar{C}_i$ over time.

### C. ORION Contact Model Construction

After running the simulation, we extracted two time series, namely the $\{C_t\}_{t \in N}$ and $\{\bar{C}_t\}_{t \in N}$, for each frequently contacted node. It was interesting to notice that all the time series were quite similar in terms of pace, even if the mobiles nodes were moving with different non-constant speeds. Figure 2 presents an example of such series.

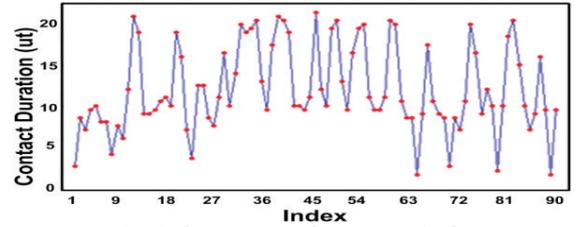

Figure 2: A chronogram for contact's duration.

The rest of this section describes the Box and Jenkins steps applied to model the proposed times series. We dubbed this model as *"ORION Contact Model"* as it is related to the targeted application scenario.

***Step 1: ORION Contact Model Identification***

We have used Minitab [22] to analyze the obtained time series. The autocorrelation function (ACF) and the partial autocorrelation function (PACF) are plotted in Figure 3 and Figure 4. According to the *ACF* and *PACF* plots, the results indicate that the best fitting model is the *ARMA(2,1)* since *PACF* presents two significant peaks (i.e. this confirms the *AR(2)* part), and the *ACF* presents one significant peak (i.e. this confirms the *MA(1)* part).

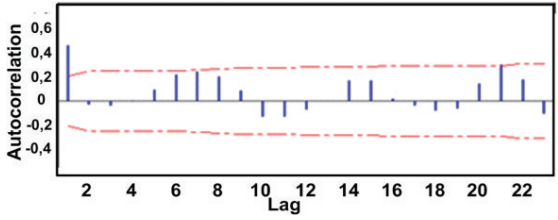

Figure 3: Autocorrelation Function (ACF) plot for contact's duration.

Based on these results, the ORION Contact Model obtained can be written as:

$$C_i = \mu + \varphi_1 C_{i-1} + \varphi_2 C_{i-2} + \theta_1 \varepsilon_{i-1} + \varepsilon_i \qquad (Eq.4)$$

Where:
- $\mu$ denotes the mean value of $C_i$
- $\varphi_1, \varphi_2, \theta_1$ denote the ORION Contact Model parameters ($\varphi_1, \varphi_2$ related to the autoregressive part and $\theta_1$ related to the moving average part).
- $\varepsilon_i, \varepsilon_{i-1}$ are assumed to be independent, identically distributed random variables sampled from a normal distribution with zero mean $\varepsilon_i \sim N(0, \sigma^2)$ where $\sigma^2$ is the variance.

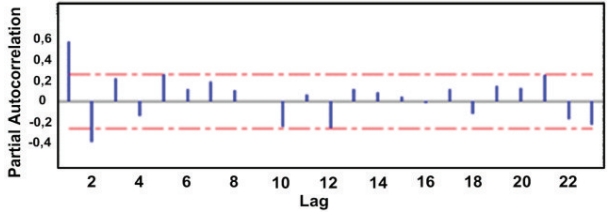

Figure 4: Partial Autocorrelation Function (PACF) plot for contact's duration.

***Step 2: ORION Contact Model Parameters Estimation***

The second step after identifying the order of the ORION Model is to estimate its parameters. For the *AR* part, the parameters can be obtained by the Yule-Walker equations [23]. The principle of the Yule-Walker equation relies on the fact that there is a direct correspondence between the parameters ($\varphi_i; i = 1 \cdots p$) and the covariance function of the process. This correspondence can be inverted to determine the parameters from the *ACF* which leads to the Yule-Walker equations:

$$\gamma_m = \sum_{k=1}^{p} \varphi_k \gamma_{m-k} + \sigma_\varepsilon^2 \delta_m$$

Where $m = 0, \ldots, p$ yielding ($p+1$) equations. $\gamma_m$ is the autocorrelation of Y. $\sigma_\varepsilon$ is the standard-deviation of the input noise process, and the $\delta_m$ is the *Kronecker Delta* function. This equation provides a way to estimate the *AR(p)* parameters by replacing the theoretical covariances with estimated values. For the *MA* part, the single parameter is obtained by identification based on the estimated *AR* parameters and the last estimation error.

***Step 3: Forecasting***

Since the orders of the ORION Model are fixed in time due to the mobility model, the computational cost of resolving linear systems can be avoided by extracting generic formulas. The node will have to compute the model parameters based on simplified mathematical expressions. Moreover, since these formulas include only aggregated data (sums, means, standard-deviations, variances …), there is no need to store all the past data; only few values are maintained at each node.

Since the data are evolving in time, we propose to compute these parameters in an incremental fashion. For two different instants $t_1 < t_2$ the estimated parameters are different because the estimation at $t_1$ takes into account the data up to $t_1$ (i.e. $[0, t_1]$) and similarly the estimation at $t_2$ takes into account all the data in $[0, t_2]$. This approach makes the estimation in real-time and more accurate with new observed data.

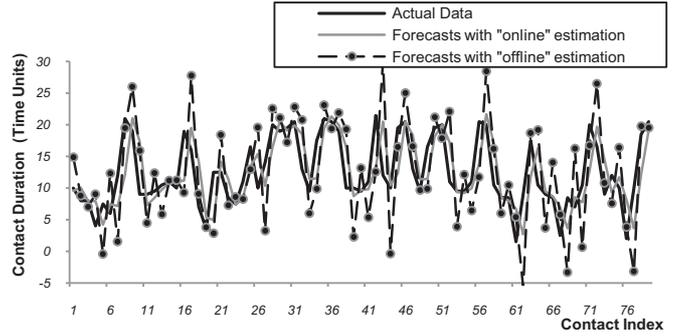

Figure 5 : Forecasting with **"online"** and **"offline"** parameter estimation.

Figure 5 presents a comparison between forecasting based on *"offline"* parameters estimation (i.e. the parameters are estimated considering the entire data set, then the future values are computed at each index using the obtained model) and forecasting based on *"online"* estimation (i.e. the parameters are estimated for each new observation, considering the available data so far). We can clearly see that our online estimation is better than the offline estimation in predicting the actual data.

### D. Forwarding Algorithm

The forwarding algorithm used in ORION is based on three criteria: (1) in order to forward a packet in a greedy manner, a node will look for the closest connected neighbor to the destination, (2) if such a node is not available, the forwarder node will look for the most advancing connected neighbor toward the destination, and finally (3) if there is no such a node, the forwarder node will schedule the data packet for the best future connected neighbor. A pseudo code of the algorithm is shown in Figure 6. A detailed version is shown in Figure 7.

| **On_Forwarding_Packet** | |
|---|---|
| 10: | nextHop ← Closest Connected Neighbor to the Destination**;** |
| 20: | **IF** (nextHop ≠ **null**) **GOTO 80**; |
| 30: | nextHop ← Most Advancing Connected Neighbor towards the Destination**;** |
| 40: | **IF** (nextHop ≠ **null**) **GOTO 80**; |
| 50: | nextHop ← Best Future Connected Neighbor**;** |
| 60: | **IF** (nextHop ≠ **null**) **GOTO 90**; |
| 70: | Store_Packet**()**; **END**. |
| 80: | Send_Packet_To**(**nextHop**)**; **END.** |
| 90: | Schedule_Packet_For**(**nextHop**)**; **END.** |

Figure 6 : Pseudo code for ORION forwarding algorithm.

| |
|---|
| *Forward_Packets ()* |
| *Packet_Queue*: a data structure where all the packets to be sent are stored. |
| *Connected_Neighbors_Set (CN)*: the set of all the neighbors that are currently in contact with this node. |
| *Estimated_Neighbors_Set (EN)*: the set of all the estimated neighbors, i.e. the nodes that we have a historical data about their contacts. |
| *Forwarder_Node (FN)*: the address of the next hop. |
| $f_{opt}$ : This function gives a score to a neighbor based on its next contact date. This function can be configured to give priority to delivery speed or delivery certainty or both of the two criteria. |
| 01: **if** (*Packet_Queue* is not Empty) { |
| 02:     *pk* = *Packet_Queue*.pop(); |
| 03:     **if** (*pk*.Next_Hop = **null**) { *//packet was not scheduled* |
| 04:        $FN = \arg Min \left\| \overrightarrow{N_{pos} pk_{dest\_pos}} \right\|, N \in CN$ |
| 05:        **if** (FN ≠ **null**) |
| 06:           send_packet(*pk*,FN); |
| 07:        **else** { *// Most advancing neighbor towards the destination* |
| 08:           $FN = \arg Max(\left\| \overrightarrow{N_{new\_pos} pk_{dest\_pos}} \right\|, \left\| \overrightarrow{N_{old\_pos} pk_{dest\_pos}} \right\|), N \in CN$ |
| 09:           **if** (FN ≠ **null**) |
| 10:             send_packet(*pk*,FN); |
| 11:           **else** { *//best future contact* |
| 12:             $FN = \arg Max\, f_{opt}(N_{next\_contact\_date}), N \in EN$ |
| 13:             *pk*.Next_Hop = *FN*; |
| 14:             *Packet_Queue*.push(*pk*); |
| 15:           } |
| 16:        } |
| 17:     } |
| 18:     **else** { *//packet was scheduled for a certain node* |
| 19:        **if** (pk.Next_Hop ∈ *CN* ) { *// is the predicted neighbor connected?!* |
| 20:           $FN = \arg Min \left\| \overrightarrow{N_{pos} pk_{dest\_pos}} \right\|, N \in CN$ |
| 21:           send_packet(*pk*, FN); |
| 22:        } |
| 23:        **else** { *// the predicted neighbor is not connected* |
| 24:           *pk*.Next_Hop = **null**; *// unscheduling the packet* |
| 25:           *Packet_Queue*.push_back(*pk*); *// storing the packet* |
| 26:           Forward_Packets(); *// Starting Over* |
| 27:        } |
| 28:     } |
| 29: } |

*Figure 7 : A detailed description of the ORION forwarding algorithm.*

## IV. PERFORMANCES EVALUATION

### A. Simulation Environment

For simulation purpose, we have considered a homogenous wireless mobile network in which nodes are randomly deployed through an area of 500m x 500m. Two nodes are selected randomly at the beginning of the simulation to act as source and destination. The source sends periodically data packets to the destination. The simulation is run for 600 seconds. To demonstrate and evaluate the performance of ORION, we used OMNeT++ 4.0 [23]. As a comparison term, we use the PRoPHET protocol. We considered variant network topologies by varying (1) the number of nodes (i.e., 30, 50, and 70 nodes) and (2) the nodes speed (i.e., 5 m/s, 10 m/s, 15 m/s and 20 m/s). For each topology, we measured various parameters: (1) the average hop count from the source to the destination, (2) the packets delivery ratio, (3) the first packet arrival, and finally (4) the average end-to-end delay. Due to space limitation, results relative to the 50 nodes topology are not shown since they are similar to those of the 70 nodes topology.

### B. Simulation Results Discussion

**a)** *Hop Count (HC):* from Figure 8 and Figure 12, we can clearly see that ORION delivers packets along fewer hops than PRoPHET and this is the case for all the three topologies and with all nodes speeds. This is achieved thanks to the twofold forwarding strategies of ORION protocol (store-and-forward and store-carry-and-forward) while PRoPHET is just a store-and-forward protocol.

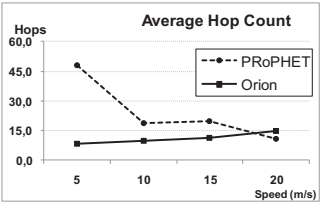

Figure 8: Average hop count in a 30 nodes topology with variant speed.

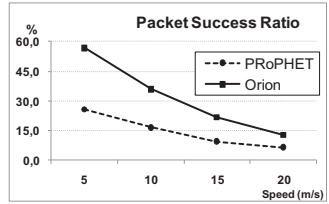

Figure 9: Packet Success Ratio in a 30 nodes topology with variant speed.

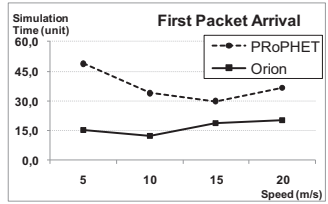

Figure 10: First Packet Arrival in a 30 nodes topology with variant speed.

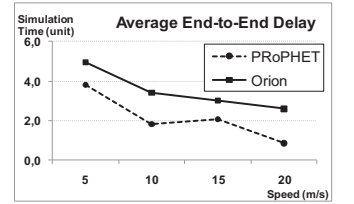

Figure 11: Average E2E Delay in a 30 nodes topology with variant speed.

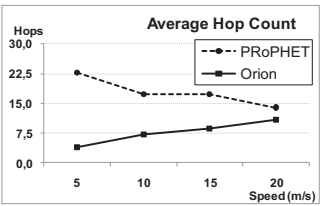

Figure 12: Average hop count in a 70 nodes topology with variant speed.

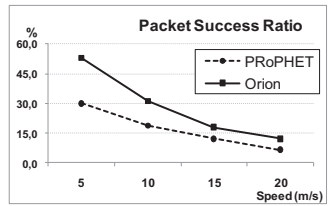

Figure 13: Packet Success Ratio in a 70 nodes topology with variant speed.

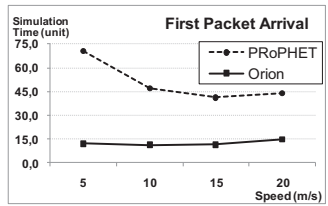

Figure 14: First Packet Arrival in a 70 nodes topology with variant speed.

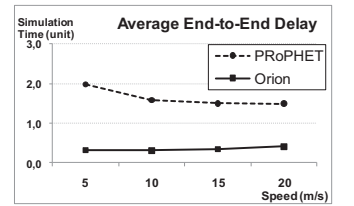

Figure 15: Average E2E Delay in a 70 nodes topology with variant speed.

**b)** *Packet Success Ratio (PSR):* Since the selection of the next forwarder node in ORION is based on three criteria (i.e. closest neighbor, most advancing neighbor, and the first future contact) rather than just one criterion (Probability of delivery success) in the case of PRoPHET, the successful node selection in ORION prevents packets from being lost; i.e., sent to nodes that cannot forward them.

This allows ORION to successfully deliver more packets than PRoPHET as shown in Figure 9 and Figure 13. From these figures, we can also notice that the impact of nodes' speed is more important in ORION than in PRoPHET. With a high speed, the accuracy of the ARMA predictions is affected since the contacts' durations will be at the same scale as prediction error margin. Thus, packets loss will be more frequent. However, the packet delivery ratio is still higher than Prophet's.

c) *First Packet Arrival (FPA) and End-to-End Transmission Delay (EED):* Because of the greedy nature of ORION, packets will always choose either the shortest or the *"earliest"* next hop making packets arrive more quickly at the destination node (Figure 10 and Figure 14) and experiencing shorter end-to-end delay (Figure 11 and Figure 15) compared to PRoPHET where the next hop is chosen based on only its success delivery probability. We have also noticed that ORION exhibits low performance compared to PRoPHET in term of end-to-end delay in a sparse topology (30 nodes). In such network, contacts are less frequent so the ARMA predictions need more time to become accurate. However with higher moving speeds, the end-to-end delay decreases.

## V. CONCLUSION

In this paper, we have described a new routing protocol, dubbed as ORION, which is suitable for mobile delay tolerant networks. Since the network dynamics are often not so random such as in city-wide inter-hotspot network interconnected through taxis, buses and vehicles, the contacts between two communicating nodes can be analyzed and, moreover, predicted. ORION routing protocol is based on greedy geographic forwarding and contacts predictions, switching between store-and-forward and store-carry-and-forward strategies in such a way, that packet forwarding is always optimal. ORION uses ARMA model online parameter estimation to predict future contacts due to its outstanding fit to this kind of network dynamics. Simulation results show that ORION routing protocol outperforms PRoPHET in terms of different metrics such as first packet arrival delay, end-to-end transmission delay, hop count, and Packet Delivery Ratio.

## REFERENCES

[1] F. Warthman, Delay-Tolerant Networks (DTNs), A Tutorial. 2003 http://www.dtnrg.org/

[2] A. Vahdat and D. Becker, "Epidemic routing for partially connected ad hoc networks," *Technical Report CS-200006*, Duke University, April 2000.

[3] Spyropoulos, T., Psounis, K., and Raghavendra, C. S. 2005. Spray and wait: an efficient routing scheme for intermittently connected mobile networks. In *Proceedings of the 2005 ACM SIGCOMM Workshop on Delay-Tolerant Networking* (Philadelphia, Pennsylvania, USA, August 26 - 26, 2005). WDTN '05. ACM, New York, NY, 252-259.

[4] A. Lindgren, A. Doria, and O. Schelen, "Probabilistic routing in intermittently connected networks," *Mobile Computing and Communications Review*, vol. 7, no. 3, July 2003.

[5] S. Jain, K. Fall and R. Patra, "Routing in a Delay Tolerant Network," in *ACM SIGCOMM*, 2004.

[6] A. Lindgren, A. Doria, and O. Schelen. "Probabilistic Routing in Intermittently Connected Networks." *Lecture Notes in Computer Science*, 3126:239 – 254, August 2004.

[7] T. Spyropoulos, K. Psounis, and C. Raghavendra. "Spray and Focus: Efficient Mobility-Assisted Routing for Heterogeneous and Correlated Mobility." *In Proc. of IEEE PerCom, 2007*.

[8] J. Burgess, B. Gallagher, D. Jensen, and B. N. Levine. "MaxProp: Routing for Vehicle-Based Disruption-Tolerant Networking." *In Proc. of IEEE INFOCOM, 2006*.

[9] A. Balasubramanian, B. N. Levine, and A. Venkataramani. "DTN Routing as a Resource Allocation Problem." *In Proc. ACM SIGCOMM, 2007*.

[10] Basu, S., Mukherjee, A., and Klivansky, S. "Time series models for internet traffic". In *INFOCOM* (March 1996), vol. 2, pp. 611–620.

[11] Liu, C., Wu, K., and Tsao, M., "Energy efficient information collection with the ARIMA model in wireless sensor networks". *In IEEE Global Telecommunications Conference GLOBECOM-05* (2005), vol. 5, pp. 2470–2474.

[12] Brockwell, P. J., and Davis, R. *Time Series: Theory and Methods*. Springer-Verlag, New York, 1987.

[13] Herbert, D., Modelo-Howard, G., Perez-Toro, C., and Bagchi, S., Fault tolerant arima-based aggregation of data in sensor networks. In *IEEE International Conference on Dependable Systems and Networks* (June 2007).

[14] Heinzelman, W., Chandrakasan, A., and Balakrishnan, H. *An application-specific protocol architecture for wireless microsensor networks. IEEE Transactions on Wireless Communications* 1, 4 (October 2002), 660–670.

[15] Banerjee, A., Majumder, K., Dutta, P., and Mondal, K. Implementation of the behavior and structure of the multihop mobile environment as a pushdown automata and birthand-death based statistical model. In *Mobile Computing and Networking (MOBICOMNET)* (2004), pp. 45–51.

[16] Singh, J. P. and Dutta, P. 2009. Temporal behavior analysis of mobile ad hoc network with different mobility patterns. *In Proceedings of the international Conference on Advances in Computing, Communication and Control* (Mumbai, India, January 23 - 24, 2009). ICAC3 '09. ACM, New York, NY, 696-702.

[17] Box, G. E. P., and G. M. Jenkins, Time Seris Analysis: Forecasting and Control, Holden-Day Inc., San Francisco, Calif., 1970.

[18] Alparslan, N. D., and Khosrow, S. A generalized random mobility model for wireless ad hoc networks and its analysis: One dimensional case. *IEEE/ACM Transactions on Networking* (TON) 15, 3 (June 2007), 602–615.

[19] Hyytia, E., Lassila, P., and Virtamo, J. A markovian waypoint mobility model with application to hotspot modeling. In *IEEE International Conference on Communications* (2006), vol. 3, pp. 979–986.

[20] Camp, T., Boleng, J., and Davies, V. A survey of mobility models for ad hoc network research. *Wireless Communications and Mobile Computing* (WCMC): Special issue on Mobile Ad Hoc Networking: Research, Trends and Applications 2, 5 (2002), 483–502.

[21] Samir Medjiah, Research Section, http://www.labri.fr/~medjiah/

[22] Minitab Inc, 2010. Minitab Statistical Software. Minitab Release 15. http://www.minitab.com.

[23] Priestley, M. B. 1981 Spectral analysis and time series / *M.B. Priestley Academic Press*, London; New York."OMNeTT++ Discreet Event Simulation System". *Internet*, 2010 (http://www.omnetpp.org/)